\documentclass{birkjour}

\usepackage{xcolor}
\definecolor{offwhite}{gray}{0.9}
\usepackage{ragged2e}

\usepackage{type1cm}        
\usepackage{makeidx}         
\usepackage{graphicx,mathrsfs}        
\usepackage{multicol}        
\usepackage[bottom]{footmisc}

\makeindex             

\usepackage{amsmath,amssymb,amsfonts,mathabx,enumitem,graphicx,subfigure}
\usepackage{setspace}

\newcommand{\dx}{\mathrm d}

\newcommand{\bra}[1]{\left\langle #1 \right\rvert}
\newcommand{\ket}[1]{\left\lvert #1 \right\rangle}

\newcommand{\fslash}[1]{\setbox 0 \hbox{$#1$} \rlap{\hbox to \wd 0 {\hss/\hss}} \box 0}
\newcommand{\matzero}{\setbox 0 \hbox{$\phantom{-1}$} \rlap{\hbox to \wd 0 {\hss0\hss}} \box 0}

\renewcommand{\theequation}{\arabic{equation}}

\begin{document}

    \title[]{Non-minimal Coupling from Consistency\\ Requirements for a Localization Operator}
    \author[]{Ross N. Greenwood}
    
    \email{ross.greenwood@ucsc.edu}
	
    \begin{abstract}
    Requiring both that a proposed localization-density operator be linear in quantum theory and that its expectation value be covariant leads to addition of terms in the Hamiltonian, coupling a scalar field to a connection on the vector-density bundle. This would appear to implicate spacetime geometry in the scalar's dynamics to a greater degree than what is prescribed by minimal coupling.
    \end{abstract}

    \maketitle
    
    \onehalfspacing
    
    {\singlespacing
    \noindent \parbox{10cm}{\small Essay written for the Gravity Research Foundation \\2020 Awards for Essays on Gravitation.}}
    
    \vspace{1cm}

\noindent
The task of defining a wave-function for the photon -- of the kind that emerges for massive fermionic fields in the non-relativistic limit -- is notoriously fraught with difficulties \cite{BIALYNICKIBIRULA1996245,Smith:2007we}.
Among what is sought from such a wave-function is a prescription for localizing particles of the field via a ``position measurement.''
%
It is argued in \cite{terno} that the expectation value of the Hamiltonian density $\mathcal H(\mathbf x)$ of a bosonic field in a single-particle state is proportional to a candidate probability density for localizing the particle, and that a positive operator-valued measure can be constructed in terms of a \textit{localization-density operator}.\footnote{Not to be confused with a \textit{density matrix} describing mixed states in quantum theory.}
\[ \Pi(\mathbf x) \equiv \bra{\Psi} \hat\Pi(\mathbf x) \ket{\Psi} = \bra{\Psi} \hat H^{-1/2} \hat{\mathcal H}(\mathbf x) \hat H^{-1/2} \ket{\Psi} \]
\[ \textstyle \mathrm{Prob}(X \in \Delta) \equiv \int_\Delta \Pi(\mathbf x) \, \dx^3\mathbf x \]
Here $\hat H^{-1/2}$ is an operator whose square is the inverse of the Hamiltonian.
The expectation values of the $(\mu,0)$ components of the stress-energy tensor are then used to derive the components of the localization current $\mathcal J^\mu$.
\[ \mathcal J^\mu(\mathbf x) \equiv \bra{\Psi} \hat H^{-1/2} \hat{T}^{\mu0}(\mathbf x) \hat H^{-1/2} \ket{\Psi} \]
It is noted that $T^{\mu\nu} = T^{\nu\mu}$ obeys a continuity equation as must $\mathcal J^\mu$, but does not transform as a four-vector density under coordinate transformations.

Taking for granted the soundness of using $T^{00}$ to construct a localization operator, we must from the outset resolve the conflicting transformation properties of $T^{\mu0}$ and $\mathcal J^\mu$.
When working with rectilinear coordinates on flat spacetime, this is straightforward: one index of $T^{\mu\nu}$ must be contracted with that of a future-directed vector density normal to the space-like hypersurface on which one would localize the particle; the result is the localization current scaled by the total energy integrated over that hypersurface.  The components $n^{\mu} T_{\mu\nu}$ are interpreted as a scaled flux of probability of localizing the particle in the 3-volume element to which $\mathbf { n}$ is normal, through the hypersurface on which the coordinate $x^\nu$ is constant.
    
If we are (probabilistically) able to localize a particle on a Cauchy surface belonging to one class of foliations of spacetime, then the same procedure should be valid for any foliation.
In the above scheme, information contained locally in the stress-energy tensor at a given event must pertain to localization on \emph{any} Cauchy surface intersecting that event.
Furthermore, a given Cauchy surface does not belong to a unique coordinate system. For this scheme to be valid, at a minimum it must prescribe the same probabilities for localization on a constant-time hypersurface common between two coordinate systems on which spatial coordinates coincide -- irrespective of which system is adopted for the calculation.
  

The quantity that, after quantization, becomes the above defined localiz-ation-density operator for the massless scalar field $\phi$ is
\begin{equation} \textstyle \Pi(\mathbf x) \equiv H^{-1/2} n^\mu T_{\mu\nu}  n^\nu H^{-1/2} = \tfrac12 \sum_\mu H^{-1/2} n^0 (\partial_\mu \phi)^2  n^0 H^{-1/2} + \cdots \label{pidef} \end{equation}
where we adopt Cartesian coordinates on flat spacetime, and $\mathbf n$ is a vector density of weight $\tfrac12$ normal to the 3-volume element\footnote{Each $n^\mu$ contributes a multiplicative factor that transforms with the coordinates as the square root of the Jacobian, such that $\Pi(\mathbf x) \, \dx^3\mathbf x$ is a tensor.}, assumed to have unit norm on the common hypersurface. The first step of quantization is to substitute linear operators suitable to act on a Hilbert space; then one substitutes the conjugate momentum density $\pi$, eliminating time derivatives of the field $\phi$ from the expression.
With the first step, we require the derivatives to be linear operators as well, obtaining for the time-like contribution to the sum
\begin{equation} \textstyle \hat\Pi^{(00)} = \frac12 \sum_\mu \hat H^{-1/2} n^0 \hat\phi \overset\leftarrow\partial_\mu \overset\rightarrow\partial_\mu \hat\phi n^0 \hat H^{-1/2} \label{pideflinear} \end{equation} 
Here $\hat\phi(\mathbf x)$ is a map from the state space $\mathscr H$ to the tensor product of $\mathscr H$ with a real vector space consisting of field values in the neighborhood of $\mathbf x$ of which $\dx^3\mathbf x$ is representative.
Acting with the operator $\overset\rightarrow\partial$ may be considered as the limit of a matrix multiplication, suitable to transform the vector $\hat\phi(\mathbf x) n^0 \hat H^{-1/2} \ket\Psi$.  
(The quantity $(\partial_\mu \hat\phi) n^0 \hat H^{-1/2} \ket\Psi$, on the other hand, cannot be so defined in terms of linear operations on $\hat\phi n^0 \hat H^{-1/2} \ket\Psi$.)

For a coordinate system in which constant-$\mathbf x$ future-directed worldlines converge with respect to inertial worldlines, the magnitude of ${\mathbf n}$ is shrinking with time to account for the decreasing weight of $\dx^3 \mathbf x$ in the volume measure.
This means that the left- and right-acting derivative operators hit the factors of $ n^0$ as well as $\hat\phi$ in \eqref{pideflinear}.
%
Only after distributing the derivatives can we substitute the conjugate momentum density $\hat\pi$ for $\hat\phi_{,0}$; 
the result differs from the prescription above by terms dependent on $n^0_{,\mu}$
\begin{multline} \textstyle \hat{\Pi}^{(00)}(\mathbf x) - \hat H^{-1/2} \hat {\mathcal H}(\mathbf x) \hat H^{-1/2} = \frac12 \sum_\mu \hat H^{-1/2} n^0_{,\mu} \hat\phi^2  n^0_{,\mu} \hat H^{-1/2} + {} \\ \textstyle \tfrac12 \hat H^{-1/2} \! \left(n^0 \hat\pi \hat\phi  n^0_{,0} + n^0_{,0} \hat\phi \hat\pi  n^0 + n^0 \hat\phi_{,i} \hat\phi  n^0_{,i} + n^0_{,i} \hat\phi \hat\phi_{,i}  n^0 \right) \! \hat H^{-1/2}
\label{operatordiff} \end{multline}
%
Since $\mathrm{Prob}( X \in \Delta)$ must have the same form no matter the foliation to which the Cauchy surface is considered to belong, the terms on the righthand side involving $ n^0_{,\mu}$ cannot remain in the expression for $\hat\Pi$.  
However, they must be included if $\hat\Pi$ as constructed above is to be a proper linear operator. 


It is of interest to determine if there is a condition on the metric given which the righthand side of \eqref{operatordiff} vanishes in every coordinate system.
Note that $\mathbf n$ in our setup can be expressed in terms of the vierbein or frame fields, so the highest derivatives of the metric that would appear in such a constraint are first-order -- vanishing in locally inertial coordinates.  
Since this quantity is not a tensor, however, this does not guarantee that that the quantity remains zero in every coordinate system.
Supposing that such a metric condition does not exist, the quantity \eqref{operatordiff} must be made a tensor by ensuring that only \textit{covariant} derivatives $n^\mu_{;\nu}$ (which vanish uniquely) appear.

To accomplish this we replace the partial derivatives in the stress-energy tensor with linear operators that include a coupling to a non-tensor auxiliary field: $ D_\mu \equiv I \partial_\mu + \mathcal G_\mu$.  
Let $\mathcal G_\mu$ be a connection on the (weight-$\tfrac12$) vector-density bundle, transforming with the coordinates in order that
$\langle n^\mu \tilde {T}_{\mu\nu} n^\nu \rangle$ is a scalar density of weight 1, where $\mathcal G_\mu \equiv (\mathcal G_\mu)^\alpha_{\phantom\alpha\beta}$ and 
\begin{multline} \textstyle \tilde T_{\mu\nu} \equiv \hat\phi \! \left( \mathcal G^\alpha_{\alpha\mu} + \overset\leftarrow\partial_\mu \right)\! \left( \overset\rightarrow\partial_\nu + \mathcal G^\beta_{\beta\nu} \right) \! \hat \phi - \tfrac12 \left( \hat\phi \overset\leftarrow\partial_\rho g_{\mu\nu} g^{\rho\sigma} \overset\rightarrow\partial_\sigma \hat\phi \right. + {} \\
\left. \hat\phi \overset\leftarrow\partial_{\rho} g_{\mu\beta} g^{\rho\sigma} \mathcal G^{\beta}_{\sigma\nu} \hat\phi + \hat\phi \mathcal G^\alpha_{\rho\mu} g_{\alpha\nu} g^{\rho\sigma} \overset\rightarrow\partial_{\sigma} \hat \phi + \hat\phi \mathcal G^\alpha_{\rho\mu} g_{\alpha\beta} g^{\rho\sigma} \mathcal G^{\beta}_{\sigma\nu} \hat\phi \right) \label{tildeT} \end{multline}
To the extent that an operator can be classified as a tensor or non-tensor, this operator is not a tensor -- until it is sandwiched between quantities that transform as vector-densities with weight $\tfrac12$.  Since this is the case when computing expectation values of the Hamiltonian and linear momenta,
the presence of $\mathcal G_\mu$ does not spoil the transformation properties of those quantities.

In adopting the operator \eqref{tildeT} to fix the definition of a localization-density operator, we have modified the Hamiltonian to include a coupling of scalar fields to the field $\mathcal G_\mu$.  
Since $\mathcal G_\mu$ acts as a connection on the tangent space, this amounts to a non-minimal coupling of the scalar field to spacetime geometry.
If this were the stress-energy operator for the purpose of localization, it seems doubtful that it would not play a role elsewhere.
%
A similar procedure can be performed for the vector gauge fields of the Standard Model, beginning at \eqref{pideflinear} with the stress-energy tensor for the $U(1)$ gauge field
\[ \hat T_{\mu\nu}^{\text{EM}} = \overset\leftarrow F_{\mu\alpha} g^{\alpha\beta} \overset\rightarrow F_{\nu\beta} - \tfrac14  \overset\leftarrow F\vphantom{F}^{\alpha\beta} g_{\mu\nu} \overset\rightarrow F_{\alpha\beta} \quad\text{with}\quad \overset\rightarrow F_{\mu\nu} \equiv  \overset\rightarrow\partial_\nu \hat A_\mu - \overset\rightarrow\partial_\mu \hat A_\nu \]
%
Perhaps this is an illustration of the unraveling  of the particle concept when translating between inertial and noninertial coordinate systems \cite{1974Natur,Davies_1975}.
If the prescription for localizing single particles becomes pathological after adopting noninertial coordinates, it may merely reflect that when viewed with respect to the new coordinates the system is no longer in a single-particle pure state, and so the localization prescription is no longer valid.

This argument has hinged on the interpretation of a quantity like $\hat\phi\ket\Psi$ as a vector living in a tensor product space formed from the state space and that of field values evaluated over the neighborhood of a point in our queried interval.  Acting with $\overset\rightarrow\partial_\mu$ yields a new element of that space with field values replaced by values of $\partial_\mu\phi$.  If this interpretation is flawed -- which is likely since the coordinates span an affine space rather than a vector space -- then $\eqref{pideflinear}$ and the conclusions that follow are erroneous.  In that case, we can obtain the naked partial derivative in \eqref{pidef} from the usual pointlike definition
\[ \lim_{h\to0}\tfrac1h (\hat\phi(\mathbf x + h \hat x) - \hat\phi(\mathbf x))n^0 H^{-1/2} \ket\Psi \]
with no illusions that something called a ``derivative operator'' is acting on a quantity $\hat\phi(\mathbf x) n^0 H^{-1/2} \ket\Psi$.
The distinction puts a spotlight on the importance of mathematical foundations in quantum field theory in determining what we predict (a notion that needed no further defense).

\vspace{1cm}
\noindent
\paragraph{Acknowledgements} 
I would like to acknowledge Daniel Terno for thought-provoking correspondence, and Joe Schindler for helpful questions and comments.  I am grateful to Anthony Aguirre and UCSC Physics for their support.

    \bibliography{reference}{}
    \bibliographystyle{unsrt}
    
\end{document}